\documentclass[sigconf, screen, nonacm]{templates/cidr-2026}

\usepackage{graphicx}
\usepackage[shortlabels]{enumitem}
\usepackage{hyperref}
\usepackage{multirow}
\usepackage{pgf}
\usepackage{lmodern}
\usepackage{booktabs}
\usepackage{geometry}
\usepackage{tabularx}
\usepackage{nameref}
\usepackage{cancel}
\usepackage{adjustbox}

\title{Theseus: A Distributed and Scalable GPU-Accelerated Query Processing Platform Optimized for Efficient Data Movement}

\begin{abstract}
Online analytical processing of queries on datasets in the many-terabyte range is only possible with costly distributed computing systems. 
To decrease the cost and increase the throughput, systems can leverage accelerators such as GPUs, which are now ubiquitous in the compute infrastructure. 
This introduces many challenges, the majority of which are related to when, where, and how to best move data around the system. 
We present Theseus -- a production-ready enterprise-scale distributed accelerator-native query engine designed to balance data movement, memory utilization, and computation in an accelerator-based system context. 
Specialized asynchronous control mechanisms are tightly coupled to the hardware resources for the purpose of network communication, data pre-loading, data spilling across memories and storage, and GPU compute tasks. 
The memory subsystem contains a mechanism for fixed-size page-locked host memory allocations to increase throughput and reduce memory fragmentation.
For the TPC-H benchmarks at scale factors ranging from 1k to 30k on cloud infrastructure, Theseus outperforms Databricks Photon by up to $4\times$ at cost parity.
Theseus is capable of processing all queries of the TPC-H and TPC-DS benchmarks at scale factor 100k (100 TB scale) with as few as 2 DGX A100 640GB nodes.

\end{abstract}

\author{
Felipe Aramburú\textsuperscript{1} \quad
William Malpica\textsuperscript{1} \quad
Kaouther Abrougui\textsuperscript{1} \quad
Amin Aramoon\textsuperscript{1} \quad \\
Romulo Auccapuclla\textsuperscript{1} \quad
Claude Brisson\textsuperscript{1} \quad
Matthijs Brobbel\textsuperscript{1} \quad 
Colby Farrell\textsuperscript{1} \quad \\
Pradeep Garigipati\textsuperscript{1} \quad
Joost Hoozemans\textsuperscript{1} \quad
Supun Kamburugamuve \quad
Akhil Nair\textsuperscript{1} \quad
Alexander Ocsa\textsuperscript{1} \quad 
Johan Peltenburg\textsuperscript{1} \quad 
Rubén Quesada López\textsuperscript{1} \quad
Deepak Sihag\textsuperscript{1} \quad \\
Ahmet Uyar\textsuperscript{1} \quad
Dhruv Vats\textsuperscript{1} \quad
Michael Wendt\textsuperscript{1} \quad
Jignesh M. Patel\textsuperscript{2} \quad
Rodrigo Aramburú\textsuperscript{1}
}

\begin{document}
\maketitle

\begingroup
\renewcommand\thefootnote{}
\footnotetext{\textsuperscript{1}Voltron Data}
\footnotetext{\textsuperscript{2}Carnegie Mellon University}
\endgroup





\renewcommand{\shortauthors}{Aramburú, Malpica et al.}


%
\section{Introduction}\label{sec:introduction}


Despite more than a decade of academic and commercial exploration, consensus remains elusive on GPUs for OLAP-style analytics, as most efforts celebrate raw compute speed yet overlook the cost of moving columnar datasets to and from device memory.
GPUs present their own problems in the form of reduced memory capacity and increased memory management complexity. 
If hardware-bound data operations are run sequentially (whether file I/O, on-GPU computation, spilling off GPUs over PCIe, network shuffle, etc.), the benefit from the addition of GPUs may not offset the additional cost of moving data.
To address these challenges, we introduce Theseus, a production-ready distributed query engine designed to leverage accelerators, specifically modern GPUs.
In this paper, we focus on Theseus' ability to use GPU accelerators, known to outperform CPUs in highly parallelizable OLAP tasks \cite{2020_Shanbhag_Crystal}.
The design of Theseus incorporates several experience-based insights that were gained over the course of its development:
\begin{enumerate}[
  label=Insight \Alph*,
]
    \item \label{ins:tight-control} Asynchronous control mechanisms tied to hardware interfaces, memories, and compute at a fine-grained level improve system utilization while hiding latency between these interfaces. 
    %
    \item \label{ins:assist-and-compete} As different asynchronous control mechanisms assist each other, care must be taken to prevent them from unintentionally competing for resources.
    %
    \item \label{ins:hide-compute-not-interface-and-memory} A data-centric engine benefits from abstractions that help easily orchestrate and optimize memory management and data movement.
\end{enumerate}


We contribute a description of Theseus' distributed worker design, where four control mechanisms, called executors, asynchronously execute tasks specific to certain system resources (\ref{ins:tight-control}) in a collaborative manner (\ref{ins:assist-and-compete}).
We also explain some key abstractions and optimizations that facilitate moving data between different memory tiers (GPU, Host, Storage) and mechanisms for proactively moving bytes onto and off of the GPU to ensure that GPU computation is not blocked by I/O and that resources are available to tasks that need them in accordance with \ref{ins:hide-compute-not-interface-and-memory}.
For brevity, our discussions restricts to the CUDA computational back-end of Theseus for NVIDIA GPUs (which is most mature), yet our insights are equally valid for its AMD ROCm backend and to some extent its Velox (CPU SIMD-accelerated) backend.
Collectively, the mechanisms in Theseus result in an efficient and scalable distributed GPU analytic query processing platform that can outperform state-of-the-art alternatives and adapt to multiple hardware configurations.


\section{Background}\label{sec:background}
There is a long history of research investigating GPU acceleration for analytic query processing, including~\cite{GovindarajuLWLM04, 2020_Shanbhag_Crystal, 2023_GPU_DB_characterization}.
In recent years, there has been a flurry of activity to build commercial GPU query processing platforms, and many of these efforts, like Theseus, use \texttt{libcudf}~\cite{libcudf2019} as a key component.  The \texttt{libcudf} library provides GPU implementations for various common operations such as scans, joins, aggregations, and filters. 

Sirius DB~\cite{SiriusDB2023} and Spark RAPIDS~\cite{SparkRAPIDS2020} provide distributed runtimes, while Polars~\cite{Polars2021} 
and RAPIDS cuDF~\cite{libcudf2019} for Pandas operate on a single node. 
Beyond \texttt{libcudf}‑based projects, HeavyDB~\cite{HeavyDB2018} and a recent Microsoft prototype~\cite{Wu2025Terabyte} also target GPU‑accelerated SQL analytics. 
A recent paper compares several GPU-accelerated databases~\cite{2023_GPU_DB_characterization}, and notes that the Crystal engine~\cite{2020_Shanbhag_Crystal} shows the highest performance (though it only supports the Star Schema Benchmark).
The potential of GPUs for real-world query processing was demonstrated with representative workloads in \cite{2016_IBM_DB2_BLU}.
An in-depth analysis and characterization of database systems employing GPUs is provided by \cite{2022_Rosenfeld_phd_thesis}.


Overall, there is a considerable interest in running analytic query processing on GPUs in a way that is cost-effective and scalable, and Theseus targets this need.
Theseus is inspired by BlazingSQL~\cite{BlazingSQL2020}, and follows a composable design philosophy (elaborated in the Composable Data Management System Manifesto~\cite{Pedreira2023Manifesto}).
Theseus adopts Apache Arrow's columnar memory model, allowing new standards and technologies to be integrated without re‑engineering core components.

\section{System Design}\label{sec:architecture}
%
%
A Theseus cluster has four core components: a \textit{Client}, a \textit{Gateway}, a \textit{Planner} (based on Apache Calcite), and \textit{Workers}. 
The Gateway serves as an intermediary between the client and the other components.
When the client submits a query, the planner creates the query plan, and then every worker receives the same physical execution plan with a different subset of files to scan. 
Theseus does not ingest the data it is operating on, but rather reads data directly from raw files, making it a true disaggregated compute data platform.



\subsection{Physical Plan Execution}\label{sec:[physical-plan]}
\begin{figure}[t]
  \centering
  \includegraphics[width=\linewidth]{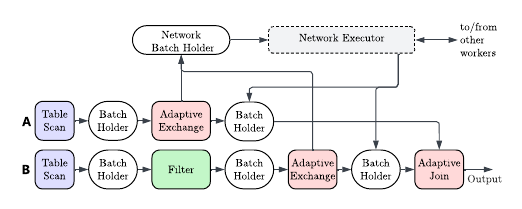}
  \vspace*{-1ex}
  \caption{Example of physical plan operators and batch holders}
  \label{fig:dag}
  \Description{Example of a physical plan's directed acyclic graph (DAG) wherein operators are connected by Batch Holders. The plan has two table scan Operators, each feeding into their own Adaptive Exchange Operators that then in turn feed an Adaptive Join Operator. The Adaptive Exchanges are seen to send batches to a Network Batch Holder which feeds a Network Executor. The Network Executor connects to and from other workers, and sends batches to the Batch Holders at the output of the Adaptive Exchange operators`.}
  \vspace*{-1em}
\end{figure}

When workers receive the physical plan generated by the planner, they create a Directed Acyclic Graph (DAG) of \textit{Operators} and \textit{Batch Holders} which is illustrated in \autoref{fig:dag}.

\textit{Operators} spawn tasks that work on a specific step of the physical query plan that are submitted to the Compute Executor (discussed in \S\ref{sec:executors}), where they are executed by leveraging GPU kernels that are invoked asynchronously from CPU threads.
Each task takes one or more \textit{batches} of input data, where a batch is a slice of all data that will flow through the operator, represented by a set of columns with the same number of rows. 
Operators ensure batches are sized according to what is suitable for GPU computation: large enough to amortize GPU kernel launch overhead and small enough to allow multiple GPU streams\footnote{GPU streams are the mechanism through which CPU threads can asynchronously launch GPU kernels. They ensure that work scheduled on the same stream is executed in order (but make no guarantee on the execution order between streams).} to run simultaneously.
Operators can schedule multiple types of tasks and can have scheduling constraints. 
Some operators, such as Filter, can schedule tasks as soon as batches arrive at their input, while others, such as Adaptive Exchange, may need to wait for a certain amount of data to arrive, as described in \autoref{sec:example_execution}.

A \textit{Batch Holder} is an abstraction of a data container that guarantees that inputs can always be stored somewhere in the system, even when the intended target memory is full (\ref{ins:hide-compute-not-interface-and-memory}).
Its data may be moved to a larger memory (including storage) when resources are scarce.
This guarantee simplifies the design of Theseus, as it encapsulates and separates the concern of \textit{where} to best move data from other control paths.
Although similar, this contrasts with CUDA Unified Memory (which allows oversubscribing to GPU memory and allowing the driver to handle spilling between Host and GPU memory) by also providing the means to move data to storage, to modify the data format (e.g., to compress it), and to explicitly move the data back to GPU memory ahead of the launch of a GPU kernel (see \autoref{sec:preload_executor}).

As shown in \autoref{fig:dag}, Batch Holders are conceptually instantiated as edges of the DAG, where data can accumulate before processing by a next operation or before being sent across the network.
Thus, each executor can operate at its own rate, and the distributed runtime is resilient to variance in the rate of its different executors during query execution.
Some operators use Batch Holders to hold data as part of their internal state, and the Network Executor uses them in its transmission buffer.

\subsection{Example Execution}\label{sec:example_execution}

\autoref{fig:dag} shows a simplified select-join DAG with a filter on Table B. 
To execute this on 4 GPUs (4 workers) using Apache Parquet files, two table scan operators start generating tasks in each worker, each task processing fractional or multiple Parquet files, depending on their size.
The tasks get pushed to the Compute Executor's queue illustrated in \autoref{fig:worker}. 

Meanwhile, the Pre-load Executor looks for scan tasks waiting in this queue, and can temporarily take ownership of the task in order to read the necessary bytes from the Parquet files into either GPU memory or host memory, depending on resource availability, ahead of computation. 
After re-inserting the scan task into the Compute Executor queue, it can continue to decompress and decode the data on the GPU. 
This eliminates the need to serialize I/O with GPU computation when parsing Parquet files (following \ref{ins:tight-control}).
Note that when the Compute Executor executes a scan task, and the bytes have not already been read from the Parquet files, it will do so itself; this way, the Pre-load Executor does not block the Compute Executor (following \ref{ins:assist-and-compete}).
As each scan task completes, its output is pushed into a Batch Holder, which may accumulate batches for the following operators, namely Adaptive Exchange A and Filter B in Figure~\ref{fig:dag}. 

An Adaptive Exchange operator exists as a pair, one for each side of a join.
A join has two phases.
First, it waits to accumulate enough input batches to estimate the total bytes it will receive, and broadcasts that information to paired Adaptive Exchange operators in all workers. 
These operators are adaptive because based on the estimates, they decide whether to hash partition or broadcast the data in the second phase during processing.
To send data to other workers, tasks utilize the Network Executor. This involves pushing batches of data along with destination information to a Batch Holder, which the Network Executor then pulls from to send the message to other workers.
The algorithm using an estimate of the data sizes to arrive instead of waiting for all the data to arrive minimizes interruption of data flow through the DAG by allowing phase two tasks to be scheduled sooner (following \ref{ins:assist-and-compete}).

In the second phase, the Adaptive Join operator must wait until some data has arrived from both Adaptive Exchange operators to schedule the data joining compute tasks.
The Compute Executor has a priority queue for tasks, designed with \ref{ins:assist-and-compete} in mind.
This priority queue is aware of the DAG.
In this example, it prioritizes the Adaptive Exchange tasks feeding into the side of the Adaptive Join that is waiting for data input.
The DAG's output can then be either written to files or retrieved from the workers by the Client.
\subsection{Executors}\label{sec:executors}
\begin{figure}[t]
  \centering
  \includegraphics[width=1\linewidth]{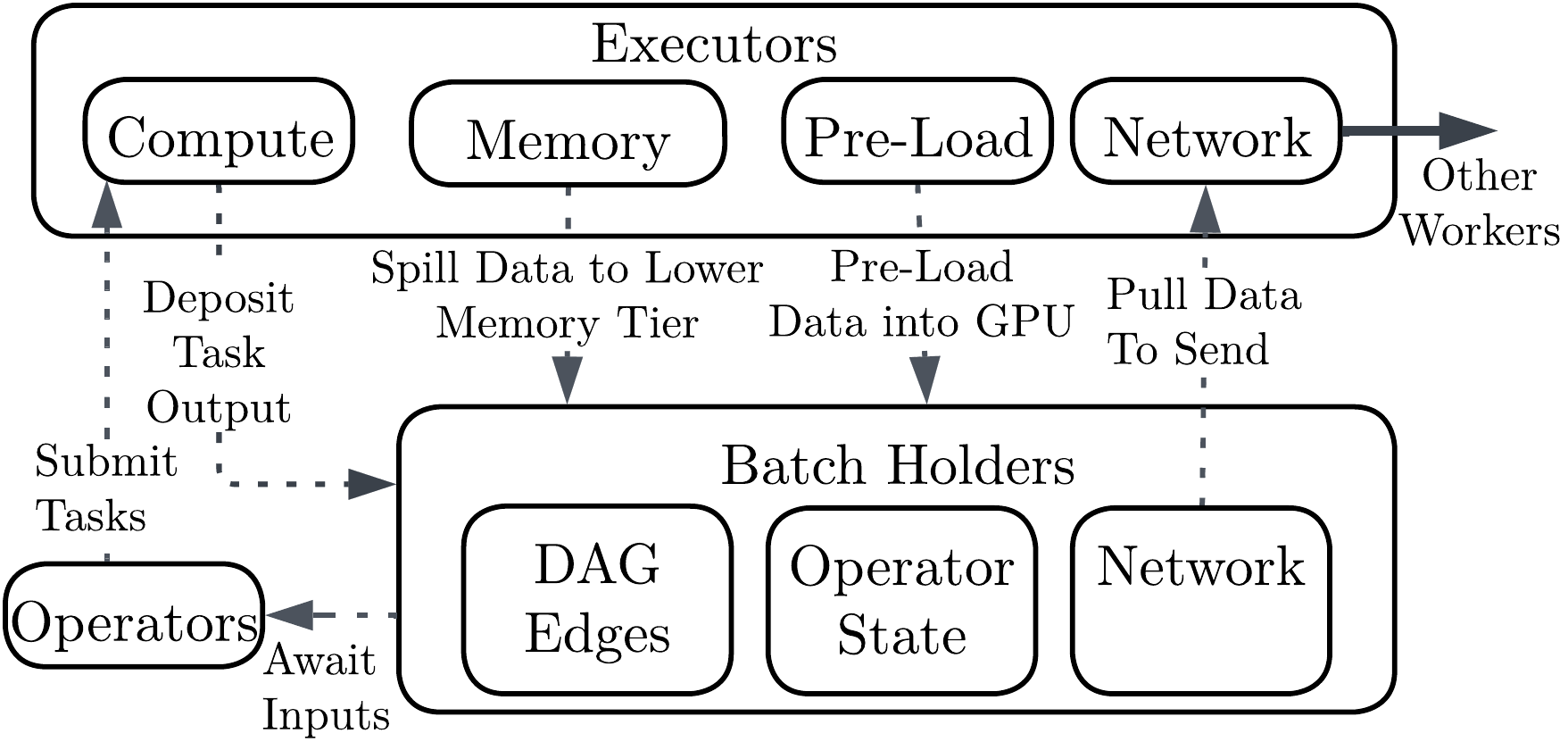}
  \caption{High-level overview of worker components}
  \label{fig:worker}
  \Description{A schematic representation of Theseus' worker internals, depicting a high-level overview of worker components, with three top-level groups of components are displayed through rectangles; Executors, Operators and Batch Holders. The executor group holds a Compute, Memory, Pre-load and Network Executor component displayed as a rectangle. The Batch Holders group contains DAG Edges, Operator State, and Network displayed as rectangles. The operators group does not display any lower-level components. The schematic contains the follow annotated arrows between components or component groups: From Operators to Compute Executor: Submit Tasks, from Compute Executor to Batch Holders: Deposit Task Output, from Memory Executor to Batch Holders: Spill Data to Lower Memory Tier, from Pre-load Executor to Batch Holders: Pre-Load Data into GPU, from Network Batch Holder to Network Executor: Pull Data to Send. An arrow is also shown indicating data flow from the Network Executor to other workers.}
  \vspace*{-1ex}
\end{figure}
\autoref{fig:worker} shows the high-level architecture of the workers. 
The components orchestrate query execution while balancing resource utilization to maximize throughput. 

Each worker process instantiates four executors: Compute, Memory, Pre-loading, and Networking. 
All executors have a number of configurable CPU threads on which they execute their tasks in parallel.
Submitted tasks are executed asynchronously.
Note that bulk computation typically happens on the GPU, so these CPU threads of these executors mainly process control flow operations described in the remained of this section.

\subsubsection{Compute Executor}\label{sec:compute_executor}
The Compute Executor executes tasks created by Operators on the GPU.
Executing a task involves several stages: reserving memory, loading input batches from batch holders into GPU memory, and finally performing the computations described by the Operator.
The Compute Executor can prioritize tasks in its queue based on different configurable schemes that can take into account a wide variety of factors, including where in the query graph the task came from and the memory tier that the input data resides in.
Each Compute Executor thread controls a separate CUDA stream using per-thread-default-stream, increasing the potential for parallel work to take place on the GPU.

\subsubsection{Memory Executor}\label{sec:executor-memory}
In order to free up GPU memory for allocations made by Compute Executor tasks, the Memory Executor runs tasks that instruct Batch Holders to spill their contents to a larger memory (e.g. from GPU memory to CPU main memory).
To decide which batches to spill, it inspects the priority queue of the Compute Executor to avoid spilling data for which compute tasks are close to being executed, which exemplifies \ref{ins:assist-and-compete}.

Before they execute, Compute Executor tasks are required to reserve (not allocate) memory with the Memory Executor.
If there is not enough memory to create a reservation, a Memory Executor task is triggered to free up the requested reservation.
These memory reservations help \textit{prevent} out-of-memory errors while compute tasks perform allocations during execution.
Each Operator keeps track of actual memory consumption of previously executed compute tasks, which feed into a heuristic that determines how much memory to reserve with the Memory Executor for the next compute task.
Compute tasks that run out of memory can be retried, improve their estimations on subsequent runs, and be divided up so that tasks are resilient to resource exhaustion and executors that can operate close to memory capacity.

As executors and batch holders operate asynchronously, situations may arise where specific memories reach capacity, which may cause reservations to induce high latency, especially when compute task make bursty reservations. 
This may dramatically slowing down query progress.
Tasked with resolving this situation before it occurs, the Memory Executor monitors all memory tiers, and if it detects that consumption reaches a threshold, it will trigger a task.
\subsubsection{Pre-loading Executor}\label{sec:preload_executor}
The Pre-loading Executor inspects the task queue of the Compute Executor (\ref{ins:assist-and-compete}).
Under configurable constraints, specific types of tasks are selected from the queue, and the Pre-loading Executor proactively initiates data transfers to ensure input data required by upcoming tasks is readily materialized.
This hides latency by eliminating the need for the Compute Executor to stall if input data is not yet materialized in GPU memory.

The Pre-loading Executor supports many modes, some of which can be enabled concurrently.
For brevity, we describe only two, whose merits are demonstrated quantitatively in \autoref{sec:results}.
In \textit{Compute Task Pre-loading} mode, input batches whose data does not yet reside in GPU memory are targeted, similar to how a CPU cache can perform prefetching (although this is not speculative).

The \textit{Byte Range Pre-loading} mode, mentioned in \autoref{sec:example_execution}, targets table scan tasks that operate on Parquet files.
File headers are retrieved first to identify the precise byte ranges required for scan operations. 
Sufficiently close byte ranges are then merged to reduce the total number of read operations.
The byte ranges are retrieved and stored directly in the task's Batch Holder in GPU or Host Memory, ensuring subsequent operations on the Compute Executor are limited to decompression and decoding. 
This approach separates storage from compute operations and allows one to maximize the data flow independently, as suggested by \ref{ins:tight-control}. 

\subsubsection{Datasource Interfaces}
Theseus implements a selection of efficient interfaces for different filesystems and storage layers.
It can leverage KvikIO, which is performant on filesytems that support Nvidia GPUDirect Storage (GDS). 
For cloud-based object stores, it can use Arrow's datasource implementations.
However, inspired by \ref{ins:tight-control}, Theseus implements a custom Object Store Datasource specifically tailored to integrate with the Byte Range Pre-loader and page-locked fixed-size buffer pool. 
It manages a pool of hot connections to object stores and coalesces multiple reads into single requests to increase throughput.
\subsubsection{Networking Executor}
The Networking Executor orchestrates sending and receiving batches over the network interface.
It can compress batches before sending with a variety of formats.
Compressing data trades computational resources and increased latency for higher network throughput, which is sensitive to the properties of the underlying network stack, as will be demonstrated in \autoref{sec:config_results}.
The network executor supports multiple back-ends, including one using TCP through the POSIX sockets API and one utilizing UCX that can leverage GPUDirect RDMA, among others.
\subsection{Host Memory Data Format}\label{sec:host_memory}
\begin{figure}[t]
  \centering
  \includegraphics[width=0.9\linewidth]{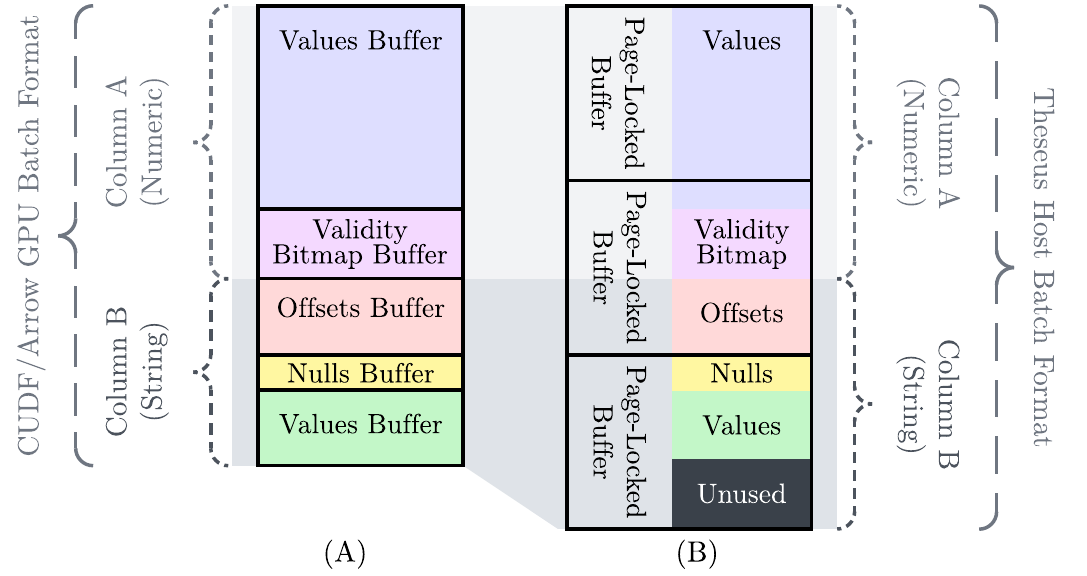}
  \caption{Example of a batch in memory (A) for CUDF/Arrow (using dynamically allocated buffers) and (B) for Theseus (using page-locked fixed-size buffers)}
  \label{fig:chunkedmemory}
  \Description{A schematic rendering of two memory layouts, A and B, where A represents the memory layout of Apache Arrow that CUDF also utilizes within GPU memory, and B how Theseus lays out data in CPU main memory. In both configurations, the memory is represented as a tall rectangle, in which smaller unfilled rectangles with thick black borders depict allocated buffers. Superimposed colored rectangles stacked on top of each other within the memory represent different pieces of Arrow-formatted data contiguously laid out in memory, implying the example is about two table columns of an Arrow RecordBatch. In both cases A and B, for the first column, there is blue values data, pink validity bitmap data. The second column has red offsets data, yellow nulls data, and green values data. In case A, the buffer rectangles exactly fit the colored rectangles, implying that the buffers are dynamically sized to match the data contents. In case B, there are three fixed size rectangles, implying the buffers are fixed-size and the data is mapped onto them. The blue values data is larger than the first fixed-size buffer, examplifying that contiguous pieces of Arrow-formatted data can spill over into the next fixed-size buffer. The third buffer of case B shows a black colored rectangle with the words "unused", implying some memory capacity overhead exists for layout B.}
\end{figure}
Batches are stored in GPU device memory using the Apache Arrow format by cuDF (\autoref{fig:chunkedmemory}A). However, in host memory, a custom memory layout utilizing page-locked memory is used to speed up data transfers between host and device memory ~\cite{nvidia-cuda-best-practices}. 
%
%
%
Large amounts of page-locked memory are slow to allocate because they require contiguous allocation and CUDA driver registration.
It cannot easily be moved like paged virtual memory, so care must be taken to prevent memory fragmentation.

To address this issue, the engine has a pool of pre-allocated fixed-size page-locked buffers which is allocated during engine initialization (\autoref{fig:chunkedmemory}B). 
Data from all columns is placed into these buffers, allowing a single column's contents to overlap multiple buffers.
This approach provides resilience to memory fragmentation at the cost of a small unused block of memory per batch.
Buffers from the same pool are also utilized as bounce buffers for network transfers and pre-loading data for table scans.
\section{Results}\label{sec:results}%
%
Multiple experiments are performed to demonstrate the performance of Theseus using the TPC-H and TPC-DS benchmark suites at various scale factors, executing the queries sequentially.
The input data used are Parquet files compressed with Zstandard with a data-page size of 1024 KB.
Row groups are dimensioned to be approximately 128 MiB. 
Decimal values are encoded with precision 11 and scale 2 using a 128-bit wide decimal type.
First, \autoref{sec:config_results} demonstrates the performance gained from the mechanisms proposed in previous sections for both on-prem and cloud-based settings.
Second, \autoref{sec:scaling_results} explores the performance and scaling behavior of Theseus on an on-prem system.
Third, \autoref{sec:photon_comparison_results} compares the performance of Theseus versus another state-of-the-art query engine on a cloud-based system.

Measurements of Theseus benchmark runs are performed using two categories of systems.
The \textbf{On-Prem} category is a GPU-accelerated cluster where each node is equipped with an Intel Xeon Platinum 8380 CPU, 4 TiB of memory, and eight NVIDIA A100-SXM4-80GB GPUs.
The nodes are connected via a 200Gb/s InfiniBand network and are connected to a high-performance 18-node WEKA distributed storage cluster which supports GPUDirect Storage and Remote Direct Memory Access (RDMA).
This is arguably a typical configuration for on-premise GPU-enabled infrastructure.
The results classified as \textbf{Cloud} were run on AWS EC2 \texttt{g6.4xlarge} instances, where each instance has 16 vCPUs, 64 GiB memory, one NVIDIA L4 GPU with 24 GiB memory and 25 Gbps peak network bandwidth. 
In all results, queries are executed from a cold start, from remote Parquet files either on WEKA storage cluster or AWS S3.

\subsection{Configuration Comparison}\label{sec:config_results}

\begin{figure}[t]
  \centering
  \begin{adjustbox}{clip,trim=0 0.03cm 0 0.002cm}
      \resizebox{\columnwidth}{!}{%
        \input{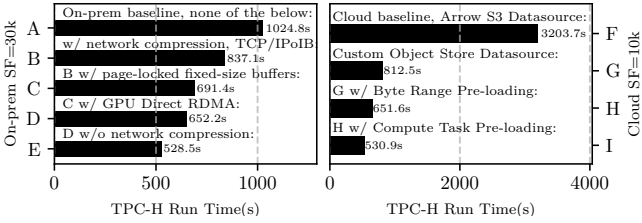}%
      }
  \end{adjustbox}
  \caption{TPC-H run time on-prem \& cloud, varying configurations}
  \label{fig:results-configs}
  \Description{A figure holding two horizontal bar graphs with different configurations on the Y-axis and TPC-H Run Time(s) on the X axis. The left-hand side shows one bar per configuration denoted with the letters A through E in alphabetical order. A descriptive text for each configuration is provided. Each bar also has a data value label. The left-hand side graph holds the following information:
  A: On-prem baseline, none of the below: 1024.8s
  B: w/ network compression, TCP/IPoIB: 837.1s
  C: B w/ page-locked fixed-size buffers: 691.4s
  D: C w/ GPU Direct RDMA: 652.2s
  E: D w/o network compression: 528.5s
  The right-hand side graph is similar to the left-hand side graph, but the Y axis appears on the right side. It holds the following information:
  F: Cloud baseline, Arrow S3 Datasource: 3203.7s
  G: Custom Object Store Datasource: 812.5s
  H: G w/ Byte Range Pre-loading: 651.6s
  I: H w/ Compute Task Pre-loading: 530.9s
  }
\end{figure}

Theseus has many configurable parameters for tuning or enabling some of its features and mechanisms to benefit specific hardware systems and queries. 
A complete exploration of these parameters is outside the scope of this work. 
\autoref{fig:results-configs} shows the results of a series of TPC-H benchmark on the on-prem system and the cloud system, with various configurations selected to demonstrate some features discussed in \autoref{sec:architecture}.
For configuration \textbf{A-E}, the run time of the TPC-H benchmark was measured at scale factor 30k using a cluster of three nodes.
Each node has 8 GPUs, thus the system is utilizing 24 GPUs.

The baseline configuration \textbf{A} uses no page-locked memory buffers or network compression, and uses the POSIX TCP API back-end which uses IPoIB on this system.
In configuration \textbf{B}, the Network Executor compresses data before it is dispatched to another worker and decompresses it when receiving, which reduces the run time by $18\%$.
In configuration \textbf{C}, the page-locked fixed-size buffer strategy from \autoref{sec:host_memory} is enabled, which results in another $17\%$ run time reduction.
In configuration \textbf{D}, GPU Direct RDMA is leveraged for worker communication, which should greatly increase the network throughput, yet only results in an overall 6\% reduction in run time.
However, because of the increased throughput capacity, resources spent on network compression were no longer best dedicated to the Network Executor.
In configuration \textbf{E} they are freed up by disabling compression, providing a final 19\% improvement.
Combined, the benefits provided by tuning Network Executor parameters and fixed-size page-locked memory constitute to a $2\times$ speedup over the baseline configuration.

\begin{figure}
  \centering
  \begin{adjustbox}{clip,trim=0 0.03cm 0 0.002cm}
  \resizebox{\columnwidth}{!}{%
    \input{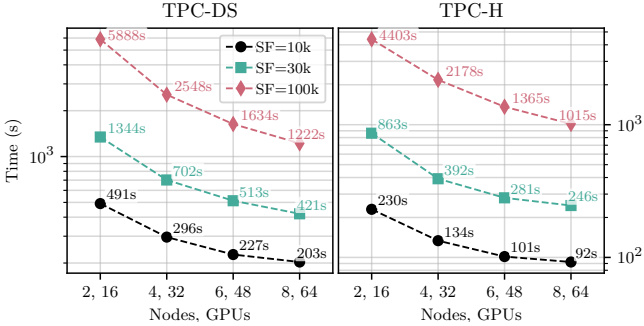}%
  }
  \end{adjustbox}
  \caption{On-prem total run time (cold queries) when scaling Theseus on TPC-DS and TPC-H at varying scale factors and node counts.}
  \label{fig:results-onprem}
  \Description{The figure shows two graphs, one for the TPC-DS benchmarks (left) and one for the TPC-H benchmarks (right). Each graph has Time (s) on the Y axis in log scale, and a node/gpu count on the X axis, where each GPU count is 8 times the node count. Each graph has three lines with four data points where the node count is 2,4,6, and 8 nodes. There is one colored line with data points using different markers for each scale factor 10k (black, circle), 30k (green, square) and 100k (red, diamond). The data presented in the graphs is as follows, formatted in the following order, benchmark, scale factor, node count, time (s):
TPC-DS, 10000, 2, 491
TPC-DS, 10000, 4, 296
TPC-DS, 10000, 6, 227
TPC-DS, 10000, 8, 203
TPC-DS, 30000, 2, 1344
TPC-DS, 30000, 4, 702
TPC-DS, 30000, 6, 513
TPC-DS, 30000, 8, 421
TPC-DS, 100000, 2, 5888
TPC-DS, 100000, 4, 2548
TPC-DS, 100000, 6, 1634
TPC-DS, 100000, 8, 1222
TPC-H, 10000, 2, 230
TPC-H, 10000, 4, 134
TPC-H, 10000, 6, 101
TPC-H, 10000, 8, 92
TPC-H, 30000, 2, 863
TPC-H, 30000, 4, 392
TPC-H, 30000, 6, 281
TPC-H, 30000, 8, 246
TPC-H, 100000, 2, 4403
TPC-H, 100000, 4, 2178
TPC-H, 100000, 6, 1365
TPC-H, 100000, 8, 1015
  }
  \vspace*{-1ex}
\end{figure}
In configurations \textbf{F-I} of \autoref{fig:results-configs}, the run time of the TPC-H benchmark was measured at scale factor 10k using a cluster of 24 machines of the Cloud instances described previously. 
Configuration \textbf{F} shows a baseline where the Arrow S3 Datasource reads Parquet files from AWS S3 with the Pre-loading Executor disabled.
In configuration \textbf{G}, Theseus uses the Custom Object Store Datasource, yielding a 75\% reduction in runtime, which illustrates the impact of \ref{ins:tight-control} where tight control around connections to S3, fixed-size allocations, and data movement yield large gains. 
In configuration \textbf{H} the Pre-Load executor's Byte Range Preloading described in \ref{sec:preload_executor} is enabled, resulting in a further 20\% reduction in runtime.
Finally, enabling the Pre-loading Executor's Task Pre-Loading (configuration \textbf{I}) reduces the runtime by another 19\%.
Both configurations H and I demonstrate the benefits of leveraging the Pre-loading Executor, highlighting how independent operation and control over a storage resource can improve throughput.

\subsection{On-prem performance and scaling behavior}\label{sec:scaling_results}
In \autoref{fig:results-onprem}, we show the total runtime for completing TPC-DS and TPC-H at scale factors 10k, 30k, and 100k with as few as two nodes (16 GPUs), up to eight nodes (64 GPUs).
Even as new GPU generations expand on-chip memory, fitting everything in GPU memory is expensive; therefore, OLAP systems at scale need to spill efficiently.

We demonstrate spilling by processing SF=100k (100TB) on two nodes, a total of 1.28 TB of GPU memory.
Note that on larger datasets (SF=100k), Theseus scales well, where four times as many GPUs provide a $4.8\times$ speedup for TPC-DS and a $4.3\times$ speedup for TPC-H. 
At SF=10k (10TB) Theseus completes TPC-H in 1.5 minutes and TPC-DS in under 4 minutes.

\subsection{Cloud performance vs. state-of-the-art}\label{sec:photon_comparison_results}
\begin{figure}[t]
  \centering
  \begin{adjustbox}{clip,trim=0 0.03cm 0 0.002cm}
      \resizebox{\columnwidth}{!}{%
        \input{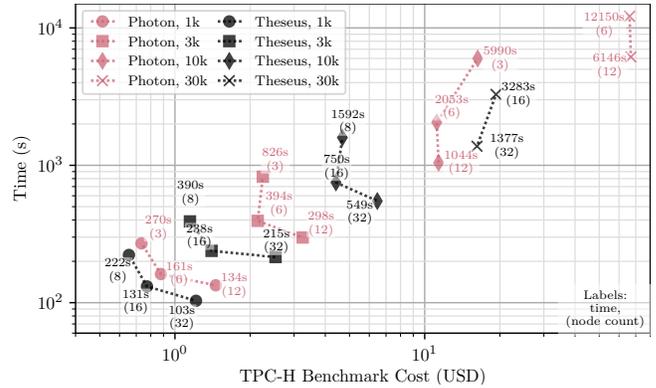}%
      }
  \end{adjustbox}
  \caption{Performance vs. cost of running TPC-H with Theseus vs. Photon on cloud clusters, varying scale factors (1k, 3k, 10k, 30k)}
  \label{fig:results-cloud-vs-photon}
  \Description{A log-log graph with time (s) on the Y-axis, and TPC-H Benchmark Cost (USD) on the X-axis. There are eight lines, half of which represent Photon and the other half representing Theseus. Each engine has a separate color, black for Theseus and red for Photon. For each engine, there is one line with a distinct marker for each scale factor: round for 1k, square for 3k, diamond for 10k, cross for 30k.
  The data presented in the graph is as follows, in the format: engine, scale factor, node count, cost, time:
Photon, 1000, 3	, 0.734790, 269.99
Photon, 1000, 6	, 0.878898, 161.47
Photon, 1000, 12, 1.456977, 133.84
Photon, 3000, 3	, 2.248507, 826.18
Photon, 3000, 6	, 2.145537, 394.17
Photon, 3000, 12, 3.246810, 298.25
Photon, 10000, 3, 16.303413, 5990.41
Photon, 10000, 6, 11.176642, 2053.33
Photon, 10000, 12, 11.365747, 1044.04
Photon, 30000, 6, 66.133698, 12149.86
Photon, 30000, 12, 66.911886, 6146.41
Theseus, 1000, 8, 0.654188, 222.48
Theseus, 1000, 16, 0.772626, 131.38
Theseus, 1000, 32, 1.214014, 103.22
Theseus, 3000, 8, 1.148248, 390.50
Theseus, 3000, 16, 1.401585, 238.33
Theseus, 3000, 32, 2.528127, 214.94
Theseus, 10000, 8, 4.682544, 1592.46
Theseus, 10000, 16, 4.412302, 750.28
Theseus, 10000, 32, 6.461579, 549.37
Theseus, 30000, 16, 19.307414, 3283.08
Theseus, 30000, 32, 16.195968, 1377.00
  }
  \vspace*{-1ex}
\end{figure}
{\small

\begin{table}[bth]
\begin{tabular}{lll|lll}
\multicolumn{3}{c|}{\textbf{Theseus}}                    & \multicolumn{3}{c}{\textbf{Photon}}                  \\
\textbf{Nodes} & \textbf{Memory} & \textbf{Cost} & \textbf{Nodes} & \textbf{Memory} & \textbf{Cost} \\ \hline
8                 & 704 GiB          & 10.59 \$/h              & 3                & 1152 GiB        & 9.80 \$/h               \\
16                & 1408 GiB         & 21.17 \$/h             & 6                 & 2304 GiB        & 19.60 \$/h              \\
32                & 2816 GiB         & 42.34 \$/h             & 12                & 4608 GiB        & 39.19 \$/h        
\end{tabular}
\caption{Cluster node count, total GPU+CPU memory \& cost}
\label{fig:cluster-size-and-costs}  
\end{table}
\vspace*{-1em}
}
\autoref{fig:results-cloud-vs-photon} shows the performance of Theseus running all queries of the TPC-H benchmark suite at scale factors (SF) 1k, 3k, 10k and 30k, compared to a state-of-the-art production-ready distributed query engine Databricks version 16.4 LTS (includes Apache Spark 3.5.2, Scala 2.13) with Photon acceleration enabled, referred to as Photon.
For Theseus, we tested on the \textbf{Cloud} configuration described previously. 
For Photon, we tested \footnote{https://github.com/voltrondata/thirdparty-benchmarks}
on AWS Graviton3 \texttt{r7gd.12xlarge} instances, where each instance has 48 vCPUs, 384 GiB memory, and 22.5 Gbps peak network bandwidth.
\autoref{fig:cluster-size-and-costs} shows costs per hour of the cluster sizes used for this experiment and how much total memory (GPU + Host) they have. 
The cluster sizes were chosen to be of similar cost over time to normalize across different instance types. 
The results show Theseus outperforming Photon at all scale factors and cluster costs. 
The smallest differential is at the smallest scale factor with the smallest cluster, with Theseus being 12.3\% faster than Photon when normalized against cost. 
In contrast, at the largest scale with the largest cluster, the difference increases to $4.46\times$ faster.
Considering that at the largest scale factor, the Databricks clusters have a 63\% higher memory capacity, they were expected to better contend at larger scale factors, but this was not demonstrated by the experiment.

\section{Additional Discussion}
While this short paper only covers a fraction of our design and implementation, in the full-length version of this paper, we plan to include additional detail and experimental results. These include an explanation of how we implement Lookahead Information Passing~\cite{LookaheadInfoPassing} in this GPU-setting to improve runtime of join-extensive queries by $\sim$50\% in some queries, how we employ Pythonic user defined functions to integrate vector search using an index-based ANN approach that leverages GPU-accelerated libraries like NVIDIA's cuVS~\cite{cuvs_github}, and additional information regarding how memory estimation, reservation, and history functions in Theseus.
 
We will also expand on ideas that did not work. 
This includes an attempt to rely on Unified Virtual Memory and driver paging, which was an order of magnitude slower than implementing our own data spilling abstractions like the Data Holder.
Dynamically allocating page-locked memory or using variable-sized pool allocators for page-locked memory was slow and led to memory fragmentation because of the diversity of the sizes of allocations. 

\section{Conclusion}\label{sec:conclusion}
This paper presents the design of Theseus, where a set of advanced asynchronous control mechanisms tightly coupled with the multitude of hardware components of a modern distributed GPU-accelerated system provide the means to fully utilize the system's capabilities in a collaborative manner.
We demonstrated how building different executors around networking, data movement, memory management, and computation enables maximizing the throughput of each executor and the system as a whole.
Proactively moving data ahead of computation, instead of reactively, whether from storage or memories into which data was spilled, is paramount to keep GPU accelerators maximally utilized.
The pooled fixed-sized page-locked buffer allocation strategy helps these mechanisms by increasing system bandwidth and avoiding memory fragmentation.
Leveraging these control mechanisms and optimizations on a contemporary cloud system, Theseus is capable of significantly outperforming state-of-the-art engines at a similar cost over time, or perform queries of similar scale at a significantly lower cost.
As the availability and capabilities of GPU accelerators in distributed computing systems worldwide rapidly increases, production-ready analytical query engines built from the ground up to leverage GPU-accelerators, such as Theseus, provide a compelling alternative to CPU-based engines.

\bibliographystyle{ACM-Reference-Format}
\bibliography{bibliography}

\end{document}